%% file: main.tex
\documentclass[conference]{IEEEtran}
\IEEEoverridecommandlockouts

\usepackage{amsmath,amssymb,amsfonts}
\usepackage{algorithm}
\usepackage{algorithmic}
\usepackage{graphicx}
\usepackage{booktabs}
\usepackage{multirow}
\usepackage{textcomp}
\usepackage{xcolor}
\usepackage{balance}
\usepackage{cite}
\usepackage{stfloats}
\usepackage{amsthm}
\begin{document}

\title{Detecting Stealthy Jamming Attacks in UAV Networks: A Bayesian Security Game Approach}

\author{
\IEEEauthorblockN{
Jianle Ba$^{\dagger}$, Shaolong Guo$^{\dagger}$, Yuntao Wang$^{\dagger}$,
Xin Wu$^{\dagger}$, Zhou Su$^{\dagger}$, and Zhendong Li$^{\ddagger}$
}
\IEEEauthorblockA{
$^{\dagger}$School of Cyber Science and Engineering, Xi'an Jiaotong University, China\\
$^{\ddagger}$School of Information and Communications Engineering, Xi'an Jiaotong University, China
}
\thanks{Corresponding Author: Zhou Su (zhousu@ieee.org).}
}

\maketitle
\input{sections/00_abstract}
\input{sections/01_section_one}

\input{sections/02_section_two}
\input{sections/03_section_three}
\input{sections/05_section_five}

\balance
\bibliographystyle{IEEEtran}
\bibliography{references}

\end{document}

%% file: sections/00_abstract.tex
\begin{abstract}
Reliable intra-swarm communication is fundamental to cooperative sensing, control, and mission execution in unmanned aerial vehicle (UAV) swarms. In open low-altitude shared-spectrum environments, coexisting UAVs may cause benign interference through spectrum reuse and signal leakage, while jamming UAVs may deliberately degrade critical intra-swarm links. Since a jamming UAV can adjust its transmit power to disguise signal-to-interference-plus-noise ratio (SINR) degradation as link fluctuation in benign interference, stealthy jammer UAV detection remains a challenge. In this paper, we propose a novel belief-driven stealthy jammer UAV detection scheme from the perspective of Bayesian security game. 
To capture stealthy interference in shared-spectrum UAV swarms, the interfering UAV is modeled as a hidden benign or jamming type, whose transmit power strategy induces the SINR observed by the defender. The defender updates its posterior belief from the observed SINR and decides whether to maintain normal operation or trigger defensive power boosting. We characterize pure-strategy perfect Bayesian equilibrium (PBE) conditions under separating structures with distinguishable type and pooling structures with indistinguishable observations.
Numerical results show that the proposed scheme improves the defender's expected utility compared to the conventional schemes.
\end{abstract}

\begin{IEEEkeywords}
UAV swarm, stealthy jamming detection, game, Bayesian inference.
\end{IEEEkeywords}

%% file: sections/01_section_one.tex
\section{Introduction}

Unmanned aerial vehicle (UAV) swarms have been widely adopted in low-altitude applications, including cooperative surveillance, emergency communications, and disaster response. 
By coordinating communication, sensing, and mobility, UAV swarms can significantly improve spatial coverage, sensing accuracy, and mission efficiency~\cite{10106022,10793113,9035635,10586873}. 
These capabilities critically rely on reliable intra-swarm wireless links, through which UAVs exchange state, control, and sensing information to support coordinated operations~\cite{10793113,9035635}. 

However, owing to spectrum reuse, the openness of air-to-air (A2A) communication, and the broadcast nature of wireless propagation, intra-swarm links are inherently vulnerable to both unintended and malicious interference from external signals \cite{10586873}. 
Such interference may arise from either benign spectrum-sharing communication or intentional jamming.
Among these factors, jamming is particularly harmful as it can significantly reduce the received signal-to-interference-plus-noise ratio (SINR), degrade the achievable data rate, delay control-message exchange, and even cause the loss of critical state information~\cite{10288199}. 
These disruptions can further impair swarm coordination, compromise mission execution, and threaten operational safety, making jamming detection a fundamental prerequisite for reliable UAV swarm communications.

Recent studies have investigated anti-jamming and interference mitigation strategies for UAV networks from several perspectives. 
Tang et al.~\cite{11203882} proposed a graph attention network-driven hierarchical learning framework for anti-jamming UAV communications, where graph attention captures complex interference \& jamming relationships and multi-agent learning enables adaptive anti-jamming policy optimization. 
For QoE-aware UAV video transmission under jamming, Xiao et al.~\cite{xiao2021uav} developed a reinforcement learning based scheme that adapts the UAV transmission strategy to maintain service quality. 
From the perspective of resource and mobility control, joint power and trajectory optimization has been studied to characterize the interaction between a UAV and a jammer through reward-region analysis~\cite{wang2023uav}. 
In air-ground integrated networks, game-theoretic modeling combined with multi-agent deep reinforcement learning has been applied to optimize UAV deployment, networking, trajectory, and power allocation against both ground and aerial jammers~\cite{10636808}. 
These studies show that learning-based adaptation, power and mobility control, and strategic optimization can help maintain link performance under interference.

Nevertheless, most existing works focus on communication performance optimization after interference occurs, or rely on the assumption that jamming attacks are explicit, persistent, or easy to distinguish from benign interference \cite{11096015}. 
In practical shared-spectrum UAV swarms, jamming can be highly stealthy and strategic \cite{10288199}. 
Instead of employing persistent high-power interference, a stealthy jammer UAV may strategically adjust its transmit power so that the resulting link degradation resembles benign interference caused by legitimate spectrum reuse. 
Meanwhile, UAV communication links inherently experience distance variation, mobility dynamics, path loss, and channel fading, all of which can induce significant SINR fluctuations.  
Therefore, the challenge is that the defender typically observes only noisy link-quality variations, while the underlying intent and type (i.e., benign or jammer) of the interfering UAV remain hidden. 
In addition, anti-jamming defensive responses incur non-negligible costs for resource-constrained UAVs: overly aggressive responses (e.g., power boosting) can lead to false-alarm losses, whereas delayed or conservative reactions may result in missed-detection losses. 

To address these challenges, we propose a Bayesian security game framework to detect stealthy jammers in UAV swarms. The key idea is to exploit the transmit-power patterns of interfering UAVs and perform belief-based inference from the resulting SINR variations to distinguish malicious jamming from benign spectrum-sharing interference. 
In the game, the interfering UAV is modeled as a strategic agent with a private type (benign or jammer), which selects its transmit power to influence the observable SINR at the defender. Based on the observed SINR, the defender updates its posterior belief regarding the attacker type and adaptively determines whether to trigger defensive actions. 
This formulation captures the coupling between adversarial behavior of the interfering UAV and belief-based response of the defender under incomplete information. 
The main contributions are as follows:
\begin{itemize}
\item We formulate the stealthy jamming detection problem in UAV swarms as a Bayesian security game, in which the attacker strategically manipulates transmit power to mimic benign interference, while the defender performs belief-based inference from SINR observations. 
\item We develop a unified payoff model that jointly captures communication performance (e.g., link rate), power-control cost, and detection-related penalties, including false-alarm and missed-detection losses. 
Based on this model, we derive the defender's belief-threshold response rule and theoretically characterize the perfect Bayesian equilibrium (PBE) of the game under separating (i.e., type revealing) and pooling (i.e., type concealing) structures.
\item Numerical simulations validate the effectiveness of the proposed scheme in enhancing defender's utility under stealthy jamming, compared with conventional schemes. 
\end{itemize}

The remainder of this paper is organized as follows. 
Section~II presents the system model. 
Section~III formulates the Bayesian security game and analyzes the game equilibrium and the belief-threshold response. 
Section~IV presents numerical results. 
Section~V concludes the paper.

%% file: sections/02_section_two.tex
\section{System Model}
The system model includes the network model, channel and power model, SINR model, and target-link rate model.
The main notations utilized in this paper are summarized in
Table~\ref{tab:notation}.

\begin{table}[t]
\centering
\caption{Key Notations}
\label{tab:notation}
\footnotesize
\begin{tabular}{c|l}
\hline
Symbol & Definition \\
\hline
$U_D(\cdot)$ & Utility of the defender UAV \\
$\theta_L$ & Legitimate UAV  type \\
$\theta_J$ & Jamming UAV type \\
$\pi_J$ & Prior probability of jamming UAV \\
$P_S^0,P_S^1$ & Baseline and boosted UAV $S$'s power \\
$\Delta P_S$ & Transmit-power increment on the target link \\
$P_L^{q}$ & Legitimate UAV power, $q\in\{\ell,h\}$ \\
$P_J^{r}$ & Jamming power, $r\in\{\ell,h\}$ \\
$g_{SD}$ & Target intra-swarm link gain \\
$g_{LD}$ & Legitimate leakage channel gain \\
$g_{JD}$ & Jamming channel gain \\
$\lambda$ & Legitimate leakage coefficient \\
$\mu(s)$ & Posterior belief that the  interfering UAV  is a jammer \\
$C_{\rm FA}$ & False-alarm loss \\
$C_{\rm MD}$ & Missed-detection loss \\
\hline
\end{tabular}
\end{table}

\subsection{Network Model}

\begin{figure}[t]
    \centering
    \includegraphics[width=0.45\textwidth]{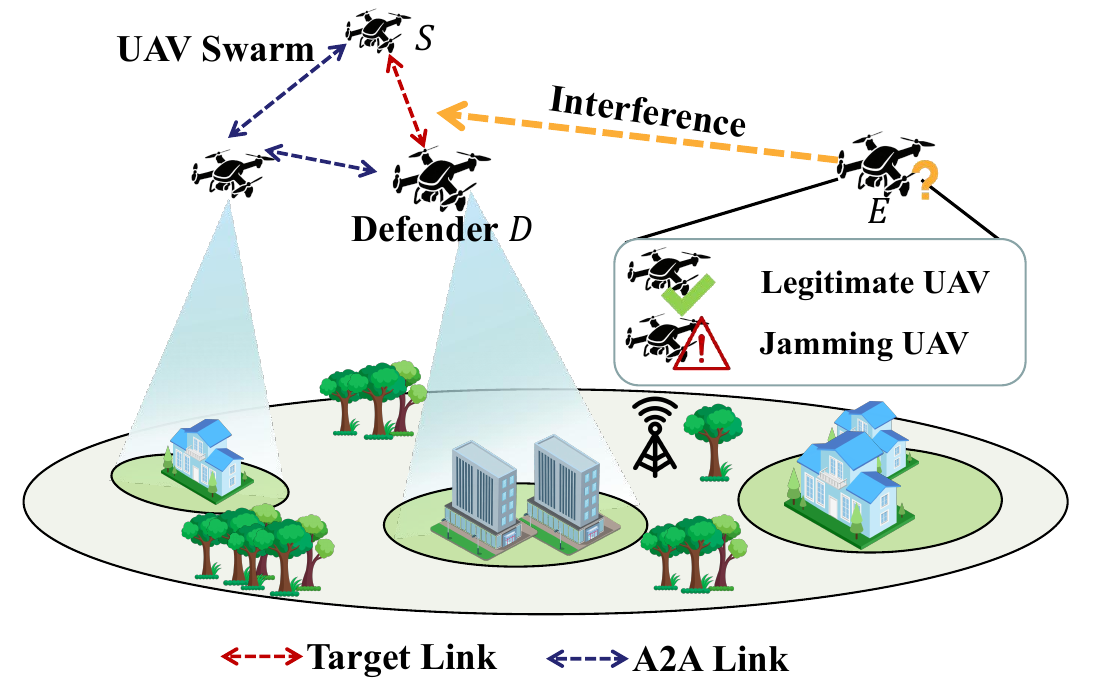}
    \caption{An illustration of UAV swarm scenario with external interference on a critical intra-swarm communication link.}
    \label{fig:scenario}
\end{figure}

Consider a UAV swarm where a UAV $S$ transmits information to a UAV $D$ over a critical intra-swarm link~\cite{9696188}. 
As illustrated in Fig.~\ref{fig:scenario}, another UAV $E$ coexists with the swarm and may interfere with the target link $S\rightarrow D$. 
In our work, UAV $D$ acts as the defender and decision-maker: it infers the link condition from the observed SINR and decides whether to request transmit-power boosting at UAV $S$.

The interfering UAV $E$ is modeled as having a private type $\theta\in\Theta=\{\theta_L,\theta_J\}$, where $\theta_L$ denotes a legitimate UAV conducting normal communication and $\theta_J$ denotes a malicious jammer. 
The legitimate UAV does not intentionally attack the UAV swarm, but its signal may cause benign interference to the target intra-swarm link due to spectrum reuse or leakage. 
In contrast, the malicious UAV intentionally transmits jamming signals to degrade the target link while attempting to remain stealthy. 
Since neither the interfering UAV's type nor its transmit power is directly observable to $D$, the defender can only rely on the resulting SINR to make its defensive decision.

\subsection{Channel and Power Model}

All A2A wireless links are modeled by a large-scale path-loss model~\cite{11096070,9159929}.
Specifically, the channel gain between UAVs $x$ and $y$ is given by $g_{xy}=G_0 d_{xy}^{-\alpha}$, where $G_0$ is the reference channel gain, $d_{xy}$ is the distance between UAVs $x$ and $y$, and $\alpha$ is the path-loss exponent.

The UAV $S$ uses a default transmit power $P_S^0$ on the target intra-swarm link. 
If the UAV $D$ triggers a defensive response, the UAV $S$ increases its transmit power to $P_S^1=P_S^0+\Delta P_S$, where $\Delta P_S>0$.
Thus, the action set of the defender UAV is $\mathcal{A}=\{a_0,a_1\}$, where $a_0$ denotes maintaining the current state, and $a_1$ denotes initiating a defensive response that triggers transmit-power boosting on the target link \cite{wang2023uav}.

The legitimate  interfering UAV  selects its communication power from $\mathcal{M}_L=\{P_L^{\ell},P_L^{h}\}$, where $0<P_L^{\ell}<P_L^{h}$.
Here, $P_L^{\ell}$ and $P_L^{h}$ denote low and high legitimate communication power, respectively. 
The jamming UAV selects its jamming power from $\mathcal{M}_J=\{P_J^{\ell},P_J^{h}\}$, where $0<P_J^{\ell}<P_J^{h}$. 
Here, $P_J^{\ell}$ and $P_J^{h}$ denote low and high jamming power, respectively. 

\subsection{Communication SINR Model}

If the interfering UAV is legitimate and selects power $P_L^{q}$, the defender UAV observes an SINR given by
\begin{equation}
    s_L^{q}
    =
    \frac{P_S^0 g_{SD}}
    {N_0+\lambda P_L^{q}g_{LD}},
    \label{eq:legitimate_sinr}
\end{equation}
where $g_{SD}$ is the channel gain of the target intra-swarm link, $g_{LD}$ is the leakage channel gain from the legitimate interfering UAV to the defender UAV, $N_0$ is the noise power, and $\lambda\in[0,1]$ is the leakage coefficient. 
When $\lambda=0$, the legitimate interfering UAV does not interfere with the target link. 
When $\lambda>0$, its communication causes unintentional interference.

If the interfering UAV is malicious and selects jamming power $P_J^{r}$, the received SINR \cite{11096015} at the defender UAV is
\begin{equation}
    s_J^{r}
    =
    \frac{P_S^0 g_{SD}}
    {N_0+P_J^{r}g_{JD}},
    \label{eq:jammer_sinr}
\end{equation}
where $g_{JD}$ is the jamming-channel gain from the jamming UAV to the defender UAV.

Since $P_L^h>P_L^\ell$ and $P_J^h>P_J^\ell$, the induced SINR values satisfy 
$s_L^\ell>s_L^h$ and $s_J^\ell>s_J^h$. The defender UAV does not directly observe either 
the type or the transmit-power level of the interfering UAV. Instead, it only observes the 
resulting SINR on the target link. Therefore, each possible type-power pair induces a 
corresponding SINR observation. Let
$\mathcal{S}=\{s_L^{\ell},s_L^{h},s_J^{\ell},s_J^{h}\}$
denote the set of possible SINR observations induced by the legitimate and malicious UAVs 
under their low and high power choices.
If two different type-power pairs induce the same SINR value, they are observationally 
indistinguishable to the defender UAV. For example, if
$s_L^{q}=s_J^{r}$, then the defender UAV cannot distinguish a legitimate 
UAV using $P_L^{q}$ from a jammer using $P_J^{r}$ based only on the observed SINR.

\subsection{Target-link Rate Model}

Let $i\in\{0,1\}$ denote the defender UAV's action index, where $i=0$ corresponds to $a_0$ and $i=1$ corresponds to $a_1$. 
After taking action $a_i$, the transmit power of UAV $S$ is $P_S^i=P_S^0+i\Delta P_S$. 
When the interfering UAV is legitimate and uses power $P_L^{q}$, the resulting target-link rate is
\begin{equation}
    R_L^{i}(q)
    =
    B\log_2
    \left(
    1+
    \frac{P_S^{i}g_{SD}}
    {N_0+\lambda P_L^{q}g_{LD}}
    \right).
    \label{eq:legitimate_downlink_rate}
\end{equation}
When the interfering UAV is malicious and adopts jamming power $P_J^{r}$, the corresponding target-link rate is
\begin{equation}
    R_J^{i}(r)
    =
    B\log_2
    \left(
    1+
    \frac{P_S^{i}g_{SD}}
    {N_0+P_J^{r}g_{JD}}
    \right),
    \label{eq:jammed_downlink_rate}
\end{equation}
where $B$ is the bandwidth of the target intra-swarm link.
We further define the interference-free reference rate of the target link as
\begin{equation}
    \bar R^{i}
    =
    B\log_2
    \left(
    1+
    \frac{P_S^{i}g_{SD}}
    {N_0}
    \right).
    \label{eq:no_jamming_rate}
\end{equation}


\section{Bayesian Security Game for Stealthy Jamming UAV Detection}


\subsection{Motivation and Game Overview}

The interaction between the defender UAV and the interfering UAV is inherently strategic under information asymmetry. The interfering UAV privately knows its type and can strategically select its transmit-power level, allowing a stealthy jammer to make its impact resemble that of a legitimate UAV. Since the defender UAV observes only the resulting SINR, it cannot directly distinguish a legitimate UAV from a malicious jammer. To capture this hidden-type uncertainty and strategic behavior, we formulate their interaction as a Bayesian security game, as follows.

\noindent\textbf{Definition 1 (Bayesian security game)}
The stealthy jamming detection problem is formulated as a game $\mathcal{G}=\left\langle \mathcal{N},\Theta,\pi_J,\sigma_E, \sigma_D,\mu,U_D,U_E \right\rangle$,
characterized by:
\begin{itemize}
    \item \textit{Players ($\mathcal{N}$):} The set is $\mathcal{N}=\{E,D\}$, where $E$ denotes the interfering UAV and $D$ denotes the defender UAV.
    
    \item \textit{Types and prior ($\Theta,\pi_J$):} The private type of interfering UAV is $\theta\in\Theta=\{\theta_L,\theta_J\}$, where $\theta_L$ and $\theta_J$ denote the legitimate and jamming types. The prior distribution is given by $\Pr(\theta=\theta_J)=1-\Pr(\theta=\theta_L)=\pi_J$.

    \item \textit{Strategies ($\sigma_E, \sigma_D$):} 
    Given its type $\theta$, the interfering UAV selects a transmit-power action $m$ according to strategy $\sigma_E(m|\theta)$, and the selected power action induces an SINR observation $s\in\mathcal{S}$ according to the SINR model in \eqref{eq:legitimate_sinr} and \eqref{eq:jammer_sinr}. The defender UAV then responds based on $s$ using the strategy $\sigma_D(a| s)$.
    

    \item \textit{Posterior belief ($\mu$):} Upon observing $s$, the defender UAV forms $\mu(s)=\Pr(\theta=\theta_J|s)$, representing its posterior probability that the interfering UAV is a jammer.
    
    \item \textit{Payoffs ($U_D,U_E$):} The utilities $U_D$ and $U_E$ characterize the payoffs of the defender UAV and the interfering UAV.
\end{itemize}

\textit{Game Process.}
Following the Harsanyi transformation \cite{11417284}, \textit{Nature} is introduced in $\mathcal{G}$ to model the uncertainty of the interfering UAV's private type.
The game proceeds as follows:
\begin{enumerate}
    \item \textit{Nature} realizes the type $\theta\in\{\theta_L,\theta_J\}$ of the interfering UAV according to the prior probability $\pi_J$.
    \item The interfering UAV privately observes its type and selects a transmit-power
    level according to $\sigma_E(m|\theta)$.
    \item The selected power induces an SINR observation $s\in\mathcal{S}$ at the defender UAV.
    \item Based on the observed SINR, the defender UAV updates its posterior belief $\mu(s)$ regarding whether the interfering UAV is malicious.
    \item The defender UAV chooses strategy $\sigma_D$, and the utilities of both players are realized.
\end{enumerate}

\subsection{Utility Model}

\subsubsection{Utility of Defender UAV}
The utility of defender UAV $D$ includes the target-link rate benefit, power-boosting cost, false-alarm loss, and missed-detection loss. 
If the interfering UAV is legitimate, the defender UAV's utility is given by
\begin{equation}
    U_D(a_i,\theta_L;q)
    =
    \omega_D R_L^{i}(q)
    -i\eta\Delta P_S
    -iC_{\rm FA},
    \label{eq:defender_utility_legitimate}
\end{equation}
where $\omega_D>0$ is the target-link rate-benefit weight, $\eta>0$ is the power-control cost coefficient, and $C_{\rm FA}>0$ is the false-alarm loss.
If the interfering UAV is malicious, its utility is
\begin{equation}
    U_D(a_i,\theta_J;r)
    =
    \omega_D R_J^{i}(r)
    -i\eta\Delta P_S
    -(1-i)C_{\rm MD},
    \label{eq:defender_utility_jammer}
\end{equation}
where $C_{\rm MD}>0$ is the missed-detection loss.


\subsubsection{Utility of Legitimate UAV}
The legitimate interfering UAV obtains a regular communication benefit, pays a transmit-power cost, and may suffer a defense-triggered penalty. 
The communication rate of the legitimate interfering UAV is
\begin{equation}
    R_V{\rm }(q)
    =
    B_L\log_2
    \left(
    1+
    \frac{P_L^{q}g_{LV}}
    {N_L}
    \right),
    \label{eq:legitimate_self_rate}
\end{equation}
where $V$ is its intended receiver UAV, $B_L$ is the bandwidth, $g_{LV}$ is the corresponding channel gain, and $N_L$ is the noise power. 
The legitimate type's utility is
\begin{equation}
    U_L(a_i,P_L^{q})
    =
    \omega_LR_{V}(q)
    -\beta_LP_L^{q}
    -i\chi_L,
    \label{eq:legitimate_utility}
\end{equation}
where $\omega_L>0$, $\beta_L>0$, and $\chi_L>0$ denote the legitimate rate-benefit coefficient, power-cost coefficient, and defense-triggered penalty, respectively.

\subsubsection{Utility of Jamming UAV}
The jamming UAV obtains attack benefit from degrading the target intra-swarm link, incurs jamming-power cost, and suffers a defense-triggered penalty. 
Its utility is expressed as
\begin{equation}
    U_J(a_i,P_J^{r})
    =
    \rho
    \left[
    \bar R^{i}-R_J^{i}(r)
    \right]
    -\beta_JP_J^{r}
    -i\delta_J,
    \label{eq:jammer_utility}
\end{equation}
where $\rho>0$ is the attack-gain coefficient, $\beta_J>0$ is the jamming power-cost coefficient, and $\delta_J>0$ is the defense-triggered penalty.

\subsection{Perfect Bayesian Equilibrium Analysis}
Since the interfering UAV privately knows its type and the defender UAV responds sequentially after observing the induced SINR, the proposed game $\mathcal{G}$ is a dynamic game with incomplete information. Accordingly, we adopt the PBE~\cite{11417284} as the solution concept.

\noindent\textbf{Definition 2 (Perfect Bayesian Equilibrium):}  
A PBE is an assessment $(\sigma_E^*,\sigma_D^*,\mu^*)$, where $\sigma_E^*$ and $\sigma_D^*$ denote the strategies of the interfering UAV and the defender UAV, and $\mu^*$ is the belief system.  
The assessment constitutes a PBE if it satisfies the following conditions: i) \emph{belief consistency}, i.e., beliefs are computed via Bayes’ rule whenever possible; ii) \emph{defender sequential rationality}, i.e., the defender UAV chooses optimal actions at each observed SINR; and iii) \emph{interfering UAV optimality}, i.e., neither type of the interfering UAV can gain from a unilateral deviation in its power selection.


In this work, we focus on pure-strategy PBE and characterize candidate equilibria as follows.

\noindent\textbf{Definition 3 (Pure-Strategy Equilibrium Types):}
For a candidate pure-strategy profile $(P_L^q,P_J^r)$, i.e., $\sigma_E(m|\theta)\in\{0,1\}$, there are two possible equilibrium types: 
i) \emph{separating equilibrium}, where two types of the interfering UAV induce distinct observations, i.e.,  $s_L^q\neq s_J^r$; 
ii) \emph{pooling equilibrium}, where they induce the same observation, i.e., $s_L^q=s_J^r$.

\subsubsection{Dynamic Belief Update}

Upon observing an on-path signal $s$\footnote{On-path signals are SINR observations produced under the candidate equilibrium strategy $\sigma_E^*$, while off-path signals are not generated by $\sigma_E^*$ and correspond to unilateral power deviations.}, the defender UAV updates its belief about the type of the interfering UAV using Bayes' rule:
\begin{equation}
    \mu^*(s)
    =
    \frac{
    \pi_J\mathbf{1}\{s_J^{r}=s\}
    }
    {
    (1-\pi_J)\mathbf{1}\{s_L^{q}=s\}
    +
    \pi_J\mathbf{1}\{s_J^{r}=s\}
    },
    \label{eq:bayes_rule}
\end{equation}
where the denominator is positive. If $s_L^q \neq s_J^r$, the defender perfectly identifies the type:
$\mu^*(s_L^q)=0$ and $\mu^*(s_J^r)=1$.
If $s_L^q = s_J^r$, the types are indistinguishable, so the posterior equals the prior:
$\mu^*(s)=\pi_J$.



\subsubsection{Optimal Response of the Defender UAV}
For a legitimate interfering UAV using power index $q$, we define
\begin{equation}
\begin{aligned}
    \Delta_L(q)
    &=
    U_D(a_1,\theta_L;q)
    -
    U_D(a_0,\theta_L;q) \\
    &=
    \omega_D
    \left[
    R_L^{1}(q)-R_L^{0}(q)
    \right]
    -\eta\Delta P_S
    -C_{\rm FA}.
\end{aligned}
\label{eq:delta_legitimate}
\end{equation}
For a malicious UAV using jamming index $r$, we define
\begin{equation}
\begin{aligned}
    \Delta_J(r)
    &=
    U_D(a_1,\theta_J;r)
    -
    U_D(a_0,\theta_J;r) \\
    &=
    \omega_D
    \left[
    R_J^{1}(r)-R_J^{0}(r)
    \right]
    -\eta\Delta P_S
    +C_{\rm MD}.
\end{aligned}
\label{eq:delta_jammer}
\end{equation}
Given posterior belief $\mu(s)$, the expected utility gain from
triggering the response is
\begin{equation}
    \Gamma(s;q,r)
    =
    [1-\mu(s)]\Delta_L(q)
    +
    \mu(s)\Delta_J(r).
    \label{eq:expected_gain}
\end{equation}

\textit{Lemma 1:} For a candidate $(P_L^{q},P_J^{r})$, the defender UAV's best response is 
\begin{equation} 
\sigma_D^*(a_1| s)=
\begin{cases}
1, & \Gamma(s;q,r)\ge 0,\\
0, & \Gamma(s;q,r)<0.
\end{cases}
\label{eq:defender_best_response} 
\end{equation}
Under $\Delta_L(q)<0<\Delta_J(r)$, this is equivalent to
\begin{equation}
\sigma_D^*(a_1| s)=1\!\!\!
\iff \!\!\!
\mu(s)\ge \mu^*(q,r)
=
\frac{-\Delta_L(q)}{\Delta_J(r)-\Delta_L(q)}.
\label{eq:belief_threshold}
\end{equation}

\textit{Proof:}
The result follows by subtracting the expected utility of $a_0$ from
that of $a_1$, which gives \eqref{eq:expected_gain}. Solving
$\Gamma(s;q,r)\ge0$ for $\mu(s)$ yields \eqref{eq:belief_threshold}.
\hfill$\blacksquare$

Let
$\Delta R_L(q)=R_L^{1}(q)-R_L^{0}(q)$ and
$\Delta R_J(r)=R_J^{1}(r)-R_J^{0}(r)$. Then
\begin{equation}
    \mu^*(q,r)
    =
    \frac{
    \eta\Delta P_S + C_{\rm FA}
    - \omega_D \Delta R_L(q)
    }
    {
    C_{\rm FA}+C_{\rm MD}
    +\omega_D\left[
    \Delta R_J(r)-\Delta R_L(q)
    \right]
    }.
    \label{eq:belief_threshold_expanded}
\end{equation}
For separating on-path signals, the best response reduces to $\sigma_D^*(a_1| s_L^q)=1 \quad \text{iff} \quad \Delta_L(q)\ge 0,$ and $\sigma_D^*(a_1| s_J^r)=1 \quad \text{iff} \quad \Delta_J(r)\ge 0.$
For pooling signals $s_L^q=s_J^r=s_{qr}$, it reduces to
$\sigma_D^*(a_1| s_{qr})=1
\quad \text{iff} \quad
\pi_J\ge \mu^*(q,r).$

\subsubsection{Optimal Selection of the Interfering UAV}
Let $\bar q$ denote the opposite power index, i.e.,
$\bar \ell=h$ and $\bar h=\ell$. Given $\sigma_D$, the legitimate and
malicious types' induced payoffs are
\begin{equation}
\Pi_L(q;\sigma_D)=\mathbb E_{a\sim\sigma_D(\cdot|s_L^q)}[U_L(a,P_L^q)],
    \label{eq:legitimate_induced_payoff}
\end{equation}
\begin{equation}
\Pi_J(r;\sigma_D)=\mathbb E_{a\sim\sigma_D(\cdot|s_J^r)}[U_J(a,P_J^r)].
    \label{eq:jammer_induced_payoff}
\end{equation}
The legitimate and malicious types have no profitable unilateral
deviations if
\begin{equation}
    \Pi_L(q;\sigma_D^*)
    \ge
    \Pi_L(\bar q;\sigma_D^*),
    \label{eq:legitimate_ic}
\end{equation}
\begin{equation}
    \Pi_J(r;\sigma_D^*)
    \ge
    \Pi_J(\bar r;\sigma_D^*).
    \label{eq:jammer_ic}
\end{equation}

\subsubsection{PBE Characterization}
Since each interfering-UAV type has two feasible power levels, we examine four pure interfering-UAV strategy candidates $(P_L^{q},P_J^{r})$. According to Definition~2, these candidates can be classified into separating and pooling equilibrium structures based on their induced SINR observations.

\textit{Proposition 1:} A candidate
$(P_L^{q},P_J^{r})$ constitutes a pure-strategy PBE if and
only if there exists a belief system $\mu^*$ and a defender UAV
strategy $\sigma_D^*$ such that: i) the on-path beliefs are Bayes
consistent, i.e.,
\begin{equation}
    \mu^*(s_L^{q})=0,\quad
    \mu^*(s_J^{r})=1,
    \quad
    \text{if } s_L^{q}\neq s_J^{r},
    \label{eq:sep_belief}
\end{equation}
\begin{equation}
    \mu^*(s_{qr})=\pi_J,
    \quad
    \text{if } s_L^{q}=s_J^{r}=s_{qr};
    \label{eq:pool_belief}
\end{equation}
ii) the defender UAV is sequentially rational at every information set according to \eqref{eq:defender_best_response}; and iii) the two types satisfy the incentive compatibility (IC) constraints \eqref{eq:legitimate_ic} and \eqref{eq:jammer_ic}.

\textit{Proof:}
Necessity follows from the PBE requirements. Bayes' rule gives
\eqref{eq:sep_belief} and \eqref{eq:pool_belief}; defender UAV
sequential rationality gives \eqref{eq:defender_best_response}; and
interfering-UAV sequential rationality gives
\eqref{eq:legitimate_ic} and \eqref{eq:jammer_ic}. Conversely, if
these conditions hold, the on-path beliefs are consistent, the
defender UAV is optimal at every information set, and neither type can
profitably deviate to its only alternative power level. Hence, the
constructed assessment is a pure-strategy PBE. \hfill$\blacksquare$

The four pure interfering UAV strategy candidates and their corresponding
row-specific IC conditions are summarized in Table~\ref{tab:pure_pbe_results}.
For each row, the defender UAV's best response is evaluated according to
\eqref{eq:defender_best_response} under the corresponding separating or
pooling belief.

\begin{table}[t]
\centering
\caption{Pure-Strategy PBE Candidates and Conditions}
\label{tab:pure_pbe_results}
\renewcommand{\arraystretch}{1.05}
\setlength{\tabcolsep}{3pt}
\scriptsize
\resizebox{\columnwidth}{!}{%
\begin{tabular}{c|c|c}
\hline
\textbf{Candidate}
& \textbf{Signal structure}
& \textbf{IC conditions} \\
\hline

$(P_L^{\ell},P_J^{\ell})$
&
\begin{tabular}{c}
$s_L^{\ell}\neq s_J^{\ell}$: separating\\
$s_L^{\ell}=s_J^{\ell}$: pooling
\end{tabular}
&
\begin{tabular}{c}
$\Pi_L(\ell;\sigma_D^*)\geq \Pi_L(h;\sigma_D^*)$\\
$\Pi_J(\ell;\sigma_D^*)\geq \Pi_J(h;\sigma_D^*)$
\end{tabular}
\\
\hline

$(P_L^{\ell},P_J^{h})$
&
\begin{tabular}{c}
$s_L^{\ell}\neq s_J^{h}$: separating\\
$s_L^{\ell}=s_J^{h}$: pooling
\end{tabular}
&
\begin{tabular}{c}
$\Pi_L(\ell;\sigma_D^*)\geq \Pi_L(h;\sigma_D^*)$\\
$\Pi_J(h;\sigma_D^*)\geq \Pi_J(\ell;\sigma_D^*)$
\end{tabular}
\\
\hline

$(P_L^{h},P_J^{\ell})$
&
\begin{tabular}{c}
$s_L^{h}\neq s_J^{\ell}$: separating\\
$s_L^{h}=s_J^{\ell}$: pooling
\end{tabular}
&
\begin{tabular}{c}
$\Pi_L(h;\sigma_D^*)\geq \Pi_L(\ell;\sigma_D^*)$\\
$\Pi_J(\ell;\sigma_D^*)\geq \Pi_J(h;\sigma_D^*)$
\end{tabular}
\\
\hline

$(P_L^{h},P_J^{h})$
&
\begin{tabular}{c}
$s_L^{h}\neq s_J^{h}$: separating\\
$s_L^{h}=s_J^{h}$: pooling
\end{tabular}
&
\begin{tabular}{c}
$\Pi_L(h;\sigma_D^*)\geq \Pi_L(\ell;\sigma_D^*)$\\
$\Pi_J(h;\sigma_D^*)\geq \Pi_J(\ell;\sigma_D^*)$
\end{tabular}
\\
\hline
\end{tabular}
}
\vspace{-2mm}
\end{table}

Let $\mathcal E$ denote the set of candidates satisfying
Proposition~1. If $\mathcal E\neq\emptyset$, each element of
$\mathcal E$ induces a pure-strategy PBE. When multiple PBEs exist, we select the defender UAV's preferred one as
\begin{equation}
    (q^\dagger,r^\dagger)
    \in
    \arg\max_{(q,r)\in\mathcal E}
    \mathbb E[U_D(q,r)],
    \label{eq:receiver_preferred_pbe}
\end{equation}
where
\begin{equation}
\begin{aligned}
\mathbb{E}[U_D(q,r)]
=
(1-\pi_J)
\sum_{i\in\{0,1\}}
\sigma_D^*(a_i| s_L^q)\,
U_D(a_i,\theta_L;q)
\\+
\pi_J
\sum_{i\in\{0,1\}}
\sigma_D^*(a_i| s_J^r)\,
U_D(a_i,\theta_J;r).
\end{aligned}
\label{eq:expected_receiver_utility}
\end{equation}
The defender-preferred rule is applied only after both interfering-UAV types satisfy the IC constraints.
If $\mathcal E=\emptyset$, then no pure-strategy PBE is selected under the current parameter setting.

%% file: sections/03_section_three.tex
\section{Numerical Results}
\label{sec:numerical_results}

This section presents numerical results to evaluate the proposed game framework. Unless otherwise specified, we set the target-link and legitimate-link bandwidths to $B=B_L=1$ MHz and the noise powers to $N_0=N_L=1.58\times 10^{-14}$~W. The UAV $S$ operates at a default transmit power level of $P_S^0=0.08$~W, and the defensive power increment is set to $\Delta P_S=0.08$~W. The low and high legitimate transmit powers are $(P_L^\ell,P_L^h)=(0.03,0.18)$~W, while the low and high jamming powers are $(P_J^\ell,P_J^h)=(0.045,0.20)$~W. The prior probability of the jamming UAV is set to $\pi_J=0.35$, and the leakage coefficient is set to $\lambda=0.30$ to represent partial interference caused by legitimate spectrum reuse. Key simulation parameters are summarized in Table~\ref{tab:simulation_parameters}. 
We compare the proposed scheme with the following benchmarks:
\begin{itemize}
    \item \emph{Always-On Power-Boosting Scheme} \cite{xiao2021uav}: This policy always triggers UAV's defensive power-boosting action.
        
    \item \emph{Fixed-Threshold Scheme}: This policy detects malicious interference UAVs by comparing the observed SINR with a baseline-calibrated threshold.

    \item \emph{Passive No-Boosting Scheme}: This policy always maintains the current state and does not trigger defensive power boosting.
\end{itemize}

\begin{table}[t]
\centering
\caption{Simulation Parameter Settings}
\label{tab:simulation_parameters}
\scriptsize
\setlength{\tabcolsep}{2.0pt}
\renewcommand{\arraystretch}{1.08}
\begin{tabular*}{\columnwidth}{@{\extracolsep{\fill}}c c|c c@{}}
\hline
Parameter & Value & Parameter & Value \\
\hline
$B$ & $1.0\,{\rm MHz}$ 
& $B_L$ & $1.0\,{\rm MHz}$ \\

$N_0$ & $1.58\times10^{-14}\,{\rm W}$ 
& $N_L$ & $1.58\times10^{-14}\,{\rm W}$ \\

$g_{SD}$ & $6.0\times10^{-12}$ 
& $g_{LD}$ & $1.5\times10^{-11}$ \\

$g_{JD}$ & $2.0\times10^{-11}$ 
& $g_{LV}$ & $3.5\times10^{-10}$ \\

$P_S^0$ & $0.08\,{\rm W}$ 
& $\Delta P_S$ & $0.08\,{\rm W}$ \\

$(P_L^{\ell},P_L^{h})$ 
& $(0.03,0.18)\,{\rm W}$ 
& $(P_J^{\ell},P_J^{h})$ 
& $(0.045,0.20)\,{\rm W}$ \\

$\omega_D$ & $2.40$ 
& $\eta$ & $24.0$ \\

$C_{\rm FA}$ & $1.20$ 
& $C_{\rm MD}$ & $6.00$ \\

$(\omega_L,\beta_L,\chi_L)$ 
& $(0.65,20.0,0.72)$ 
& $(\rho,\beta_J,\delta_J)$ 
& $(3.00,16.0,0.80)$ \\

$\pi_J$ & $0.35$ 
& $\lambda$ & $0.30$ \\

\hline
\end{tabular*}
\end{table}

\begin{figure}[t]
    \centering
    \includegraphics[width=0.5\textwidth]{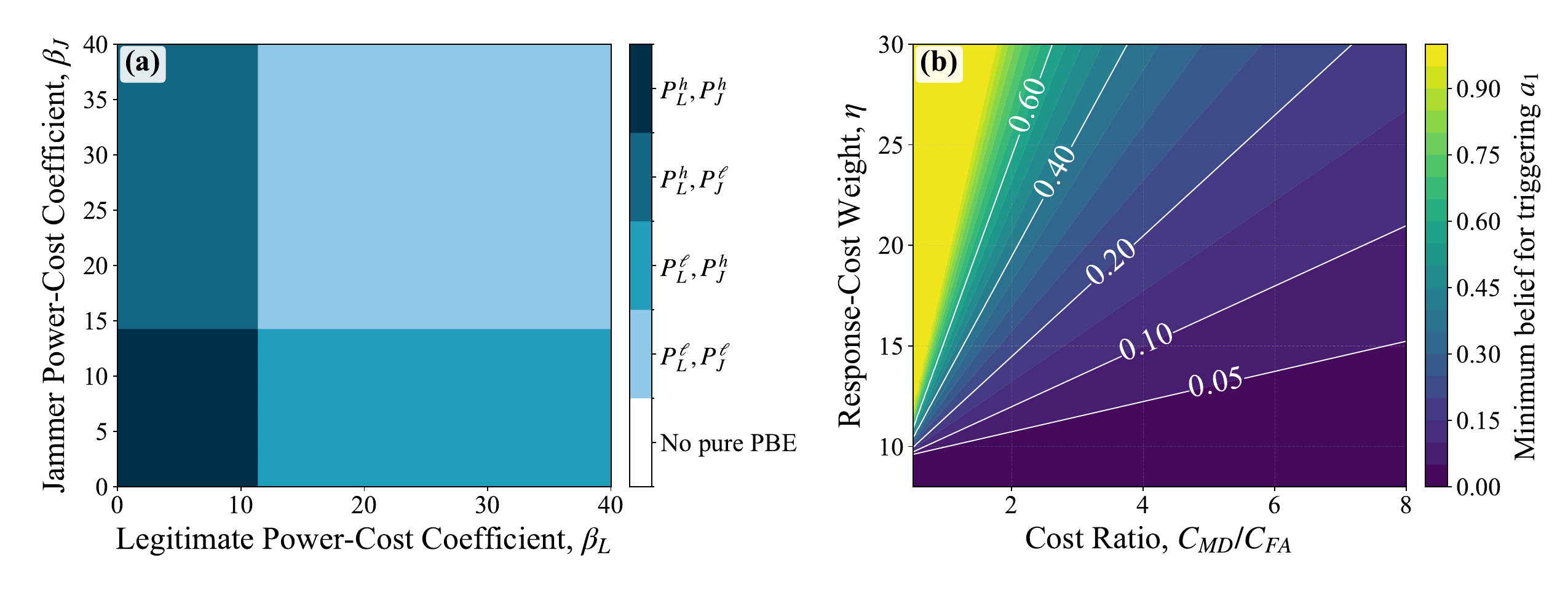}
    \caption{Equilibrium outcomes of proposed scheme: (a) defender UAV preferred pure-PBE regions; and (b) defender UAV response threshold.}
    \label{fig:equilibrium_mechanism}
\end{figure}

Fig.~\ref{fig:equilibrium_mechanism} illustrates the equilibrium mechanism of the proposed Bayesian security game. Fig.~\ref{fig:equilibrium_mechanism}(a) presents the defender-preferred pure-strategy PBE regions under different power costs, revealing how equilibrium outcomes vary with legitimate and malicious power incentives. Fig.~\ref{fig:equilibrium_mechanism}(b) shows that a higher response-cost weight raises the belief threshold for triggering defensive response $a_1$, whereas a higher missed-detection cost lowers it.

\begin{figure}[t]
    \centering
    \includegraphics[width=0.34\textwidth]{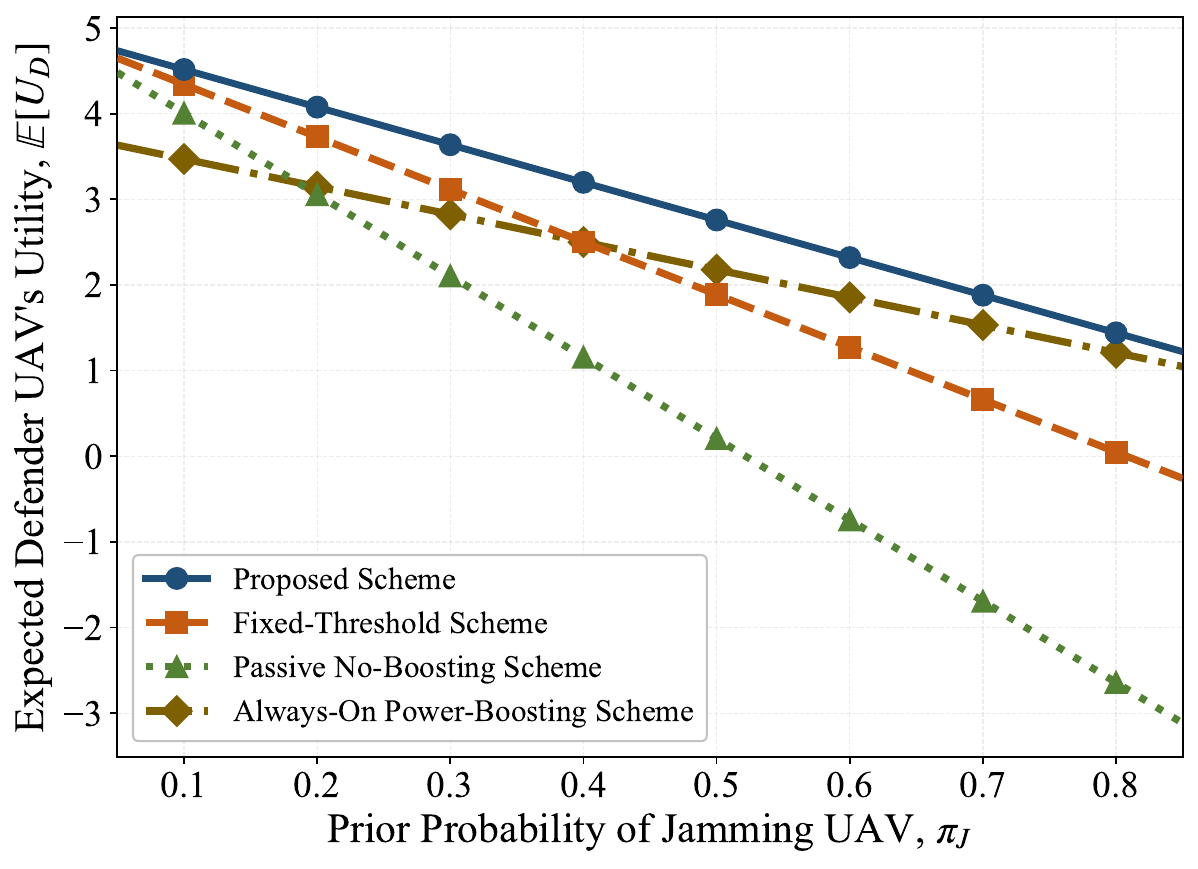}
    \caption{Expected defender UAV’s utility versus the prior probability of the jamming UAV.}
    \label{fig:defender_utility_prior}
\end{figure}

Fig.~\ref{fig:defender_utility_prior} compares the expected defender utility under different prior probabilities of a jamming type. As this prior varies, the proposed PBE-based scheme consistently outperforms the fixed-threshold, passive no-boosting, and always-on power-boosting baselines. The gain stems from jointly modeling the interfering UAV’s type-dependent power selection, defender UAV’s SINR-based posterior belief update, and its belief-threshold defensive power-boosting rule.

\begin{figure}[htbp]
    \centering
    \includegraphics[width=0.34\textwidth]{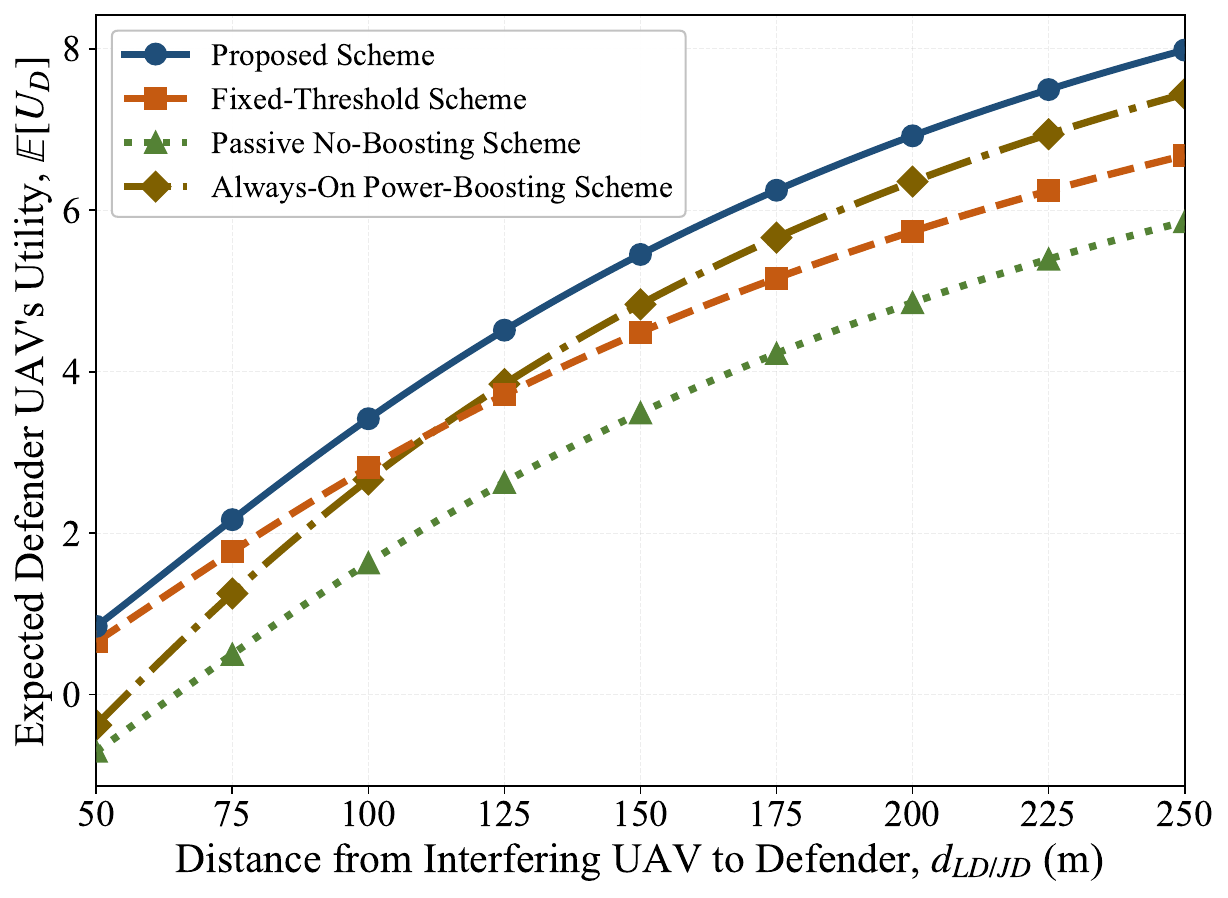}
    \caption{Expected defender UAV’s utility versus the  distance from interfering UAV to defender.}
    \label{fig:defender_utility_distance}
\end{figure}

Fig.~\ref{fig:defender_utility_distance} shows the defender UAV's expected utility as the distance between the interfering UAV and the defender UAV varies. The proposed PBE-based scheme consistently outperforms the benchmarks over the considered distance range. This result indicates that the proposed belief-driven detection and response mechanism remains effective under distance-dependent interference attenuation, in which the received interference strength and the induced SINR vary with the interfering UAV's position.

%% file: sections/05_section_five.tex
\section{Conclusion}

In this paper, we proposed a Bayesian security game framework for stealthy jamming detection in UAV swarms. The proposed framework exploits transmit-power-induced SINR variations as informative signals to distinguish stealthy jamming from benign spectrum-sharing interference, enabling the defender UAV to update its belief about the interference type and make adaptive defensive decisions. By jointly considering link-rate protection, defensive power cost, false-alarm loss, and missed-detection loss, the framework provides a cost-aware detection-and-response mechanism that improves the reliability of critical intra-swarm communications while avoiding unnecessary power boosting. Numerical results validated the equilibrium behavior and showed that the proposed scheme improves the expected defender UAV’s utility over conventional schemes. 
For future work, we will extend the framework to scenarios with multiple interferers and time-varying channels.

%% file: references.bib
@ARTICLE{10106022,
  author={Wang, Yuntao and Su, Zhou and Xu, Qichao and Li, Ruidong and Luan, Tom H. and Wang, Pinghui},
  journal={IEEE/ACM Transactions on Networking}, 
  title={A Secure and Intelligent Data Sharing Scheme for {UAV}-Assisted Disaster Rescue}, 
  year={2023},
  month={Dec.},
  volume={31},
  number={6},
  pages={2422-2438},
  doi={10.1109/TNET.2022.3226458}}

@ARTICLE{10793113,
  author={Ceviz, Ozlem and Sen, Sevil and Sadioglu, Pinar},
  journal={IEEE Communications Surveys \& Tutorials}, 
  title={A Survey of Security in UAVs and {FANETs}: {Issues}, Threats, Analysis of Attacks, and Solutions}, 
  year={2025},
  month={Oct.},
  volume={27},
  number={5},
  pages={3227-3265},
  doi={10.1109/COMST.2024.3515051}}

@ARTICLE{9035635,
  author={Su, Zhou and Wang, Yuntao and Xu, Qichao and Zhang, Ning},
  journal={IEEE Transactions on Dependable and Secure Computing}, 
  title={{LVBS}: {Lightweight} Vehicular Blockchain for Secure Data Sharing in Disaster Rescue}, 
  year={2022},
  month={Jan.--Feb.},
  volume={19},
  number={1},
  pages={19-32},
  doi={10.1109/TDSC.2020.2980255}}

@ARTICLE{10586873,
  author={Wu, Qirui and Zhang, Yirun and Yang, Zhaohui and Shikh-Bahaei, Mohammad R.},
  journal={IEEE Transactions on Wireless Communications}, 
  title={Deep Learning for Secure {UAV} Swarm Communication Under Malicious Attacks}, 
  year={2024},
  month={Oct.},
  volume={23},
  number={10},
  pages={14879-14894},
  doi={10.1109/TWC.2024.3419923}}

@ARTICLE{10288199,
  author={Ye, Runlong and Peng, Yuyang and Al-Hazemi, Fawaz and Boutaba, Raouf},
  journal={IEEE Transactions on Communications}, 
  title={A Robust Cooperative Jamming Scheme for Secure {UAV} Communication via Intelligent Reflecting Surface}, 
  year={2024},
  month={Feb.},
  volume={72},
  number={2},
  pages={1005-1019},
  doi={10.1109/TCOMM.2023.3325488}}

@ARTICLE{11203882,
  author={Tang, Xiao and Zhao, Kexin and Shen, Chao and Lin, Chenhao and Liu, Shuai and Wang, Bohui and Niyato, Dusit and Han, Zhu},
  journal={IEEE Transactions on Wireless Communications}, 
  title={Graph Attention Network-Driven Hierarchical Learning for Anti-Jamming {UAV} Communications}, 
  year={2026},
  volume={25},
  number={},
  pages={5432-5445},
  doi={10.1109/TWC.2025.3618614}}

@article{xiao2021uav,
  author={L. Xiao and Y. Ding and J. Huang and S. Liu and Y. Tang and H. Dai},
  title={{UAV} Anti-Jamming Video Transmissions With {QoE} Guarantee: A Reinforcement Learning-Based Approach},
  journal={IEEE Transactions on Communications},
  volume={69},
  number={9},
  pages={5933--5947},
  year={2021},
  month={Sep.},
  doi={10.1109/TCOMM.2021.3087787}
}

@article{wang2023uav,
  author={H. Wang and G. Ding and J. Chen and Y. Zou and F. Gao},
  title={{UAV} Anti-Jamming Communications With Power and Mobility Control},
  journal={IEEE Transactions on Wireless Communications},
  volume={22},
  number={7},
  pages={4729--4744},
  year={2023},
  month={Jul.},
  doi={10.1109/TWC.2022.3228265}
}

@ARTICLE{10636808,
  author={Liao, Chengjian and Xu, Kui and Hu, Guojie and Xia, Xiaochen and Wei, Chen and Xie, Wei and Li, Chunguo and Wang, Yurong},
  journal={IEEE Transactions on Vehicular Technology}, 
  title={Game Theory and Multi–Agent {DRL} Based Anti-Jamming Transmission for Integrated Air-Ground Network}, 
  year={2024},
  month={Dec.},
  volume={73},
  number={12},
  pages={19565-19581},
  doi={10.1109/TVT.2024.3443263}}

@ARTICLE{11096015,
  author={Wen, Yingkun and Wang, Fengshuan and Wang, Hui-Ming and Li, Junhuai and Qian, Jin and Wang, Kan and Wang, Huaijun},
  journal={IEEE Transactions on Wireless Communications}, 
  title={Joint Trajectory and Power Design With Cooperative Jamming {UAV} Assistance Based on Reinforcement Learning}, 
  year={2026},
  volume={25},
  number={},
  pages={1258-1271},
  doi={10.1109/TWC.2025.3590170}}

@ARTICLE{9696188,
  author={Wang, Yuntao and Chen, Weiwei and Luan, Tom H. and Su, Zhou and Xu, Qichao and Li, Ruidong and Chen, Nan},
  journal={IEEE/ACM Transactions on Networking}, 
  title={Task Offloading for Post-Disaster Rescue in Unmanned Aerial Vehicles Networks}, 
  year={2022},
  month={Aug.},
  volume={30},
  number={4},
  pages={1525-1539},
  doi={10.1109/TNET.2022.3140796}}

@ARTICLE{11096070,
  author={Qi, Nan and Wang, Shuqi and Wu, Daolong and Zhang, Lei and Jia, Luliang and Tian, Chang and Zhan, Ming},
  journal={IEEE Internet of Things Journal}, 
  title={Game-Theoretic Learning-Enabled Multi-{UGV} Fairness-Aware and Timely Data Collection in Industrial {WSNs}}, 
  year={2025},
  month={Dec.},
  volume={12},
  number={23},
  pages={49036-49050},
  doi={10.1109/JIOT.2025.3592231}}

@ARTICLE{9159929,
  author={Wang, Yuntao and Su, Zhou and Zhang, Ning and Benslimane, Abderrahim},
  journal={IEEE Transactions on Network Science and Engineering}, 
  title={Learning in the Air: {Secure} Federated Learning for {UAV}-Assisted Crowdsensing}, 
  year={2021},
  month={Apr.--Jun.},
  volume={8},
  number={2},
  pages={1055-1069},
  doi={10.1109/TNSE.2020.3014385}}

@ARTICLE{11417284,
  author={Guo, Shaolong and Wang, Yuntao and Su, Zhou and Pan, Yanghe and Luan, Tom H. and Luo, Xizhao},
  journal={IEEE Transactions on Networking}, 
  title={{FLET}: Game-Theoretic Free-Riding Mitigation via Test Tasks in Federated Learning}, 
  year={2026},
  volume={34},
  number={},
  month={Feb.},
  pages={3995-4010}}
